# Giant intrinsic dichroism in *β*-Ga$_2$O$_3$ enables filter-free, high-fidelity polarization division multiplexing

*Yonghui Zhang\*, Rui Zhu, Huili Liang, Guochao Zhao, Shuli Wei, Qing Lu\*, Zengxia Mei\**

Yonghui Zhang, Guochao Zhao, Shuli Wei, Qing Lu
School of Physics and Optoelectronic Engineering, Shandong University of Technology, 255000 Zibo, Shandong, P. R. China.
Emails: yhzhang@sdut.edu.cn, luqing198909@163.com

Yonghui Zhang, Rui Zhu, Huili Liang, Zengxia Mei
Dongguan Institute of Materials Science and Technology, Chinese Academy of Sciences, 523808 Dongguan, Guangdong, P. R. China.
Emails: zxmei@iphy.ac.cn

Funding: National Natural Science Foundation of China (Grant No. 12504060, 12574218, 52572167) and Shandong Provincial Natural Science Foundation (Grant No. ZR2025MS1030).

Conventional polarization detection relies on external filters, which incur significant efficiency loss and polarization crosstalk, especially in the deep ultraviolet band where subwavelength nanofabrication is challenging. Here, we report that monoclinic *β*-Ga$_2$O$_3$ exhibits intrinsic giant polarization dichroism, allowing near-ideal polarization photodetection without external optical elements and coherent polarization-division multiplexing (PDM) capability. The giant dichroism originates from the crystallographic symmetry-driven selectivity of optical transitions, which, combined with a large valence band splitting, results in vastly distinct absorption for orthogonal polarizations. A theoretical analysis of the transition selection rules in *β*-Ga$_2$O$_3$ reveals only the ***E**//**c***-polarized *vb$_1$*-to-conduction band transition is activated, within the 245–258 nm spectral window. An admirable polarization ratio surpassing 500 and a polarization crosstalk ratio below 0.2% is hence achieved. The polarization-sensitive photodetector exhibits a high responsivity of 73 A/W and fast response (~20 ms). Furthermore, we showcase its practical utility in PDM free-space communication, successfully decoding encoded optical signals, and demonstrate its capability for high-fidelity Stokes vector retrieval. The intrinsic anisotropy of *β*-Ga$_2$O$_3$, dictated by its crystal symmetry, lays the groundwork for filter-free, high-fidelity polarization polarimetry. This work further paves the way for a general design principle in next-generation optoelectronics that harness polarization transition selection rules.

1. Introduction

Polarimetry reveals an unseen aspect of light—polarization—a realm invisible to naked eyes, which carries a wealth of information beyond the intensity.[1–5] Detection of light's polarization state will reveal materials' textures and stress distributions etc., which favors catching the camouflaged targets in defense,[6,7] providing diagnostic contrast in biomedical imaging,[8] and visualizing internal stress in materials.[9,10] More significantly, it can be utilized for polarization division multiplexing (PDM) in optical communications,[11] which can expand the information capacity to meet the growing transmission demands with no need for additional physical channels. The success of the most advanced polarimetry systems to date stems from the use of subwavelength grids (SWGs) in infrared polarization-sensitive detectors,[12] which offer high integrability and reliability—attributes that have been consistently proven in practice.[10] However, this advantage does not translate to deep-ultraviolet (DUV) band, where the short wavelengths cause the difficulty and cost of fabricating SWGs to soar.[13] This limitation creates a critical technological gap. This fundamental constraint necessitates a paradigm shift away from metasurface filters like SWGs and toward intrinsic materials with innate DUV polarization dichroism.[14–16]

$\beta$-Ga$_2$O$_3$, with a quasi-direct bandgap of 4.83 eV and anisotropic monoclinic crystal structure, possesses a native sensitivity in the DUV spectrum, making it ideal for direct DUV polarimetry applications.[17–23] Furthermore, its high application value in polarimetry stems directly from an ultra-high polarization ratio (PR) that exceeds other materials by orders of magnitude, a desirable feature for PDM technology.[20,24–26] **Figure 1** presents a comparative analysis of PR values of photodetectors based on various anisotropic materials. $\beta$-Ga$_2$O$_3$ exhibits an ultrahigh PR exceeding $10^3$, significantly outperforming other polarization-sensitive materials, including black phosphorus (bP)[27] and its homo[28]/heterojunctions.[3,29]; perovskite-based materials;[30,31] and other materials such as h-BN[32] and transition metal dichalcogenides (TMDs).[33–35] A low PR indicates significant polarization crosstalk, which degrades detection fidelity by introducing noise and measurement inaccuracies. This occurs as the undesired response to the orthogonal polarization obscures the true signal, ultimately compromising the reliability of polarization-sensitive applications. However, although there are numerous reports on $\beta$-Ga$_2$O$_3$-based polarization detection, most current studies focus primarily on achieving record-high device performance. For high-precision optical communication applications such as PDM, polarization crosstalk can significantly degrade signal accuracy. Therefore, it is crucial to fundamentally understand the operational mechanism of $\beta$-Ga$_2$O$_3$ polarization detectors and mitigate the adverse effects of polarization crosstalk for their practical deployment.

In this context, $\beta$-Ga$_2$O$_3$ emerges as a premier material candidate. Its intrinsically ultrahigh PR, which originates from its unique crystal symmetry and remarkable polarization dichroism properties, effectively minimizing such crosstalk. The superior polarization detection performance of $\beta$-Ga$_2$O$_3$, and the working mechanism, will be systematically elucidated via group theory analysis, theoretical calculations, and photodetection experiments.

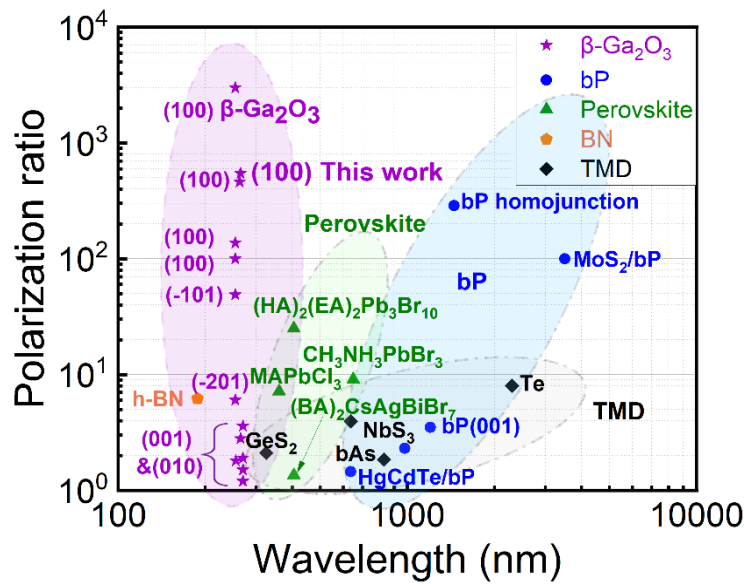

**Figure 1.** Statistical data of PR in polarization-sensitive photodetectors with diverse material systems.

## 2. Group-theoretical analysis of transition selection rules in $\beta$-Ga$_2$O$_3$

The exceptional polarization photodetection performance of $\beta$-Ga$_2$O$_3$ is achieved through its three following intrinsic properties.

## 2.1 High photo-to-dark current ratio (PDCR)

The high PDCR in *β*-Ga₂O₃ results from an ultralow dark current ($I_{dark}$) ,[36] which stems from a low intrinsic carrier concentration (*n*). This is ascribed to both its ultrawide bandgap, which thermodynamically suppresses carrier generation, and its intrinsic properties of C2/m space group, wherein the non-degenerate nature of the conduction band minimum (CBM) at the Γ point yields a lower effective density of states ($N_c$) and consequently a reduced *n*. This is a critical foundation for achieving a high PR, because PR, defined as the ratio of photocurrents under two orthogonal polarization states, is fundamentally constrained by the device's inherent signal-to-noise ceiling, which is set by the PDCR.

## 2.2 Large valence band (VB) splitting energy ($\Delta E_{vb}$)

The large splitting energy minimizes the overlap between the absorption edges, thereby giving rise to well-separated response bands for orthogonal polarization excitations. *β*-Ga₂O₃ crystallizes in a monoclinic structure with low symmetry, which results in pronounced anisotropy in its optoelectronic properties. Belonging to the C2/m space group, its symmetry operations include a two-fold rotation axis ($C_2$) along the ***b***-axis and a mirror plane ($\sigma_h$) perpendicular to it, as shown in **Figure 2**(a). These symmetries govern the polarization-dependent transition selection rules by imposing constraints on the symmetry of its electronic wavefunctions, thereby forming the physical basis for polarization photodetection. As reported in our recent work,[20] the VB of *β*-Ga₂O₃ splits into a manifold of six sub bands, indexed $vb_1$ to $vb_6$, under the influence of crystal-field splitting and spin-orbit coupling. The splitting between these levels is considerable, specifically $\Delta E_{vb1-6} > 0.5$ eV. Interband optical transitions in *β*-Ga₂O₃ are governed by two fundamental selection rules: the parity-selection rule and the polarization-selection rule.

### 2.2.1 Parity-selection rule

The CBM at Γ, with its Ga-4*s* orbital character, is fully symmetric and of even parity (irreducibly represented as $\Gamma_1^+$ in Figure 2(b)). The CBM at Γ point is with full symmetry ($\Gamma_1^+$), while the $vb_1$, $vb_2$, $vb_4$, bands are irreducibly represented as $\Gamma_2^-$, the $vb_6$ bands are irreducibly represented as $\Gamma_1^-$. Details of the irreducible representations for $\beta$-$Ga_2O_3$ can be found in Supporting information S1. The transition probability is governed by the dipole matrix element $M_{fi}$:

$$M_{fi} = \mathbf{E} \cdot \langle \psi_f | \hat{d} | \psi_i \rangle \quad (1)$$

where $\psi_i$ and $\psi_f$ are the wavefunctions of the initial and final states, respectively, $\hat{d} = -e \cdot \hat{r}$ is the electric dipole operator and $\mathbf{E}$ is the electric field. A transition is dipole-allowed if and only if the transition matrix element $M_{fi} \neq 0$ upon integration over the entire space, which requires that $M_{fi}$ possesses even parity as following

$$\Pi(\psi_f) \times \Pi(\hat{d}) \times \Pi(\psi_f) = +1 \quad (2)$$

where $\Pi(\psi_f)$, $\Pi(\hat{d})$ and $\Pi(\psi_i)$ are their parity quantum numbers. Given that $\psi_f$ is even (+1) and $\hat{d}$ is odd (-1), the parity of the initial state must be odd (-1). This foundational constraint is why electronic transitions originating from the even-parity $vb_3$ and $vb_5$ states to the CBM are symmetry-forbidden.

### *2.2.2 Polarization-selection rule*

The matrix element $M_{fi}$ in equation (1) implies that interband transitions in $\beta$-$Ga_2O_3$ require a non-zero projection of $\mathbf{E}$ onto the transition dipole moment ($\mu_{fi} = \langle \psi_f | \hat{d} | \psi_i \rangle$) direction. To investigate the polarization-selection rules, density functional theory (DFT) calculations were employed to determine the orbital-projected character of the relevant bands. As shown in Figure 2(c), at the $\Gamma$ point, the CBM is dominated by the isotropic Ga 4$s$ orbital. In contrast, the valence bands $vb_1$, $vb_2$, $vb_4$, and $vb_6$ exhibit anisotropic O 2$p$ orbital character. Specifically, $vb_1$ is constituted by $pz$ and $px$ orbitals and dominated by the $pz$ component, whereas $vb_2$ exhibits strong $px$ character. $vb_4$ is likewise dominated by the $px$ orbital. Meanwhile, $vb_6$ is predominantly characterized by contribution from the $py$ orbital.[37,38] Note that ***xyz*** represent the Cartesian coordinate axes in the laboratory frame, while ***abc*** denote the axes of the monoclinic crystal coordinate system. The correspondence between them is defined as: ***x*** ⊥ ***bc***-plane, ***y*** ∥ ***b***, and ***z*** ∥ ***c***, as can be seen in Figure 2(d). Transitions from these valence bands to the CBM can be excited when the electric field vector of the polarized light has a nonzero projection onto the major lobe direction of the respective *p* component.[39]

*For transitions from the $\Gamma_2^-$ valence bands (vb₁, vb₂) to the $\Gamma_1^+$ CBM:* The direct product of the $\psi_f$ and $\psi_i$ is $\Gamma_1^+ \otimes \Gamma_2^- = \Gamma_2^-$. The dipole transition requires that the irreducible representation of $\hat{d}$ be contained in the direct product of the representations of the final and initial states [Equation (3)].

$$\Gamma(\hat{d}) \subset \Gamma_1^+ \otimes \Gamma_2^- = \Gamma_2^- \tag{3}$$

In the case of *β*-Ga₂O₃, the dipole operator components $\hat{d}_a$ and $\hat{d}_c$ transform as the irreducible representation $\Gamma_2^-$ (see Supporting information S1 for details). This symmetry match confirms that the transition dipole moment $\mu_{fi}$ is confined to the ***ac***-plane. Consequently, the corresponding interband transition can only be excited by light polarized within this plane, i.e., for ***E***∥***a*** or ***E***∥***c***. The responses to ***E***∥***a*** and ***E***∥***c*** polarizations are fundamentally linked and cannot be decoupled, a comprehensive discussion about which can be found in Supporting information S2.

*For transitions from the vb₄ valence band to CBM:* Although the transition from the valence band *vb₄* to the conduction band *cb* is very weak,[37] it may still influence polarization-sensitive photoresponse. Even though the *vb₄* band is mainly formed by 2*px* orbitals in this coordination, its transition to the *cb* strongly depends on optical excitation with ***E***//***b***.[38] This behavior may not rely entirely on electric dipole transitions but could be associated with practical imperfections. The signal might originate from out-of-plane electric field components due to the high numerical aperture of the microscope objective, which could excite mixed transitions, or from a potentially enhanced dissociation rate compared to other excitons[40].

*For transitions from $\Gamma_1^-$ valence band (vb₆) to $\Gamma_1^+$ CBM:* The irreducible representation of $\hat{d}$ must be contained in the direct product of the irreducible representations in Equation (4).

$$\Gamma(\hat{d}) \subset \Gamma_1^+ \otimes \Gamma_1^- = \Gamma_1^- \tag{4}$$

In other words, the dipole transition requires the $\hat{d}$ operator with $\Gamma_1^-$ symmetry. This condition is fulfilled by the $\hat{d}_b$ component, making the transition dipole-allowed specifically for ***E*∥*b*** polarization.

The group theory analysis establishes the fundamental symmetry framework, determining which transitions are allowed and their general polarization constraints. The transition selection rules, combined with large valence band splitting energy *ΔE*$_{vb}$, create a broad selective absorption window, as schematically illustrated in Figure 2(e), where giant dichroism is expected between the no- and indiscriminate- absorption regime. Consequently, photodetectors based on *β*-Ga$_2$O$_3$ are expected to exhibit a polarization-sensitive band, within which, the detector responds only to one linear polarization of light while remaining blind to the orthogonal ones. This implies that the operational wavelength must be carefully selected to minimize polarization crosstalk and achieve a high polarization ratio. Therefore, identifying the optimal wavelength for this purpose becomes a critical step.

As schematically shown in Figure 2(f), the operational wavelength for polarization photodetection should be restricted to wavelengths longer than 230 nm. Shorter wavelengths correspond to photon energies sufficient to excite electrons from multiple valence bands, thereby compromising polarization selectivity. Secondly, operating within the 245–258 nm band is optimal for enhancing the contrast between ***E*//*c*** and ***E*//*b*** responses, as it selectively excites ***E*//*c*** transitions while suppressing those from ***E*//*b*** polarizations. The large energy splitting over 0.5 eV between *vb*$_1$ and *vb*$_6$ provides robust spectral isolation, which is paramount for minimizing polarization crosstalk and achieving a high PR.

**2.3 Intrinsic decoupling of *b*-axis from the *ac*-plane**

The decoupling of ***b***-axis from the ***ac***-plane is key to eliminating polarization crosstalk. To achieve a high PR, DUV light in the 245-258 nm range should be incident normal to the (100) plane (denoted as ***a*\*** direction). In this geometry, when the incident light is polarized along the ***E*//*c*** direction [Figure 2(g)], the resultant ***E*** inside the crystal possesses only *E*$_c$ component, with no contributions along both ***a*** and ***b*** axes, thereby maximizing the photoresponse for *E*//*c* polarization. Conversely, if the incident light is polarized along the *E*//*b* direction [Figure 2(h)], neither of the considered interband transition would be excited, resulting in a minimal response.

It is crucial to note that polarization crosstalk between the *E*//*a* and *E*//*c* channels cannot be completely eliminated, even under ideal excitation conditions. This residual crosstalk originates from two fundamental physical properties: Firstly, β-Ga$_2$O$_3$ is a biaxial crystal. Unless light propagates exactly along one of its optical axes, it decomposes into ordinary (o) and extraordinary (e) waves with orthogonal polarizations inside the crystal. This decomposition effectively rotates the polarization direction of the internal electric field relative to the incident polarization, as schematically illustrated in [Figure 2(i)]. Secondly, the monoclinic *β*-angle (≠ 90°) inherently introduces off-diagonal components (e.g., $\varepsilon_{xz}$) in the dielectric tensor. Consequently, an incident *E$_c$* induces an electric displacement vector along the *a*-axis (*D$_a$*). This displaced charge, in turn, generates a weak but non-zero internal *E$_a$* field component via the inverse dielectric tensor transformation. The cross-coupling effect is discussed in detail in supporting information S2.

The unique low-symmetry crystal structure of *β*-Ga$_2$O$_3$ endows it with giant intrinsic anisotropy, making it a compelling material platform for polarimetry. We have systematically elucidated the physical mechanism behind achieving an ultrahigh PR in *β*-Ga$_2$O$_3$-based detectors. We identify the optimal operational parameters: a detection window of 245-258 nm, which exclusively targets the *vb$_1$*-related transition, and a crystal orientation with illumination normal to the (100) plane. This specific configuration maximizes the response to *E*//*c* polarization while effectively suppressing the crosstalk from *E*//*a* and *E*//*b*. Guided by this design principle, we fabricated a polarization detector from a (100)-oriented *β*-Ga$_2$O$_3$ single crystal, which exhibited exceptional performance. Polarization-resolved photoresponse spectra revealed well-separated cutoff edges, yielding a high PR of 547 at 265 nm under normal incidence to the (100) plane. The practical value of this polarization-sensitive photodetector was demonstrated across key applications, ranging from DUV PDM communication, where suppressed crosstalk enhanced decoding accuracy and bandwidth potential, to precision polarimetry, where the exceptional PR allowed for accurate retrieval of Stokes parameters and linear polarization characteristics.

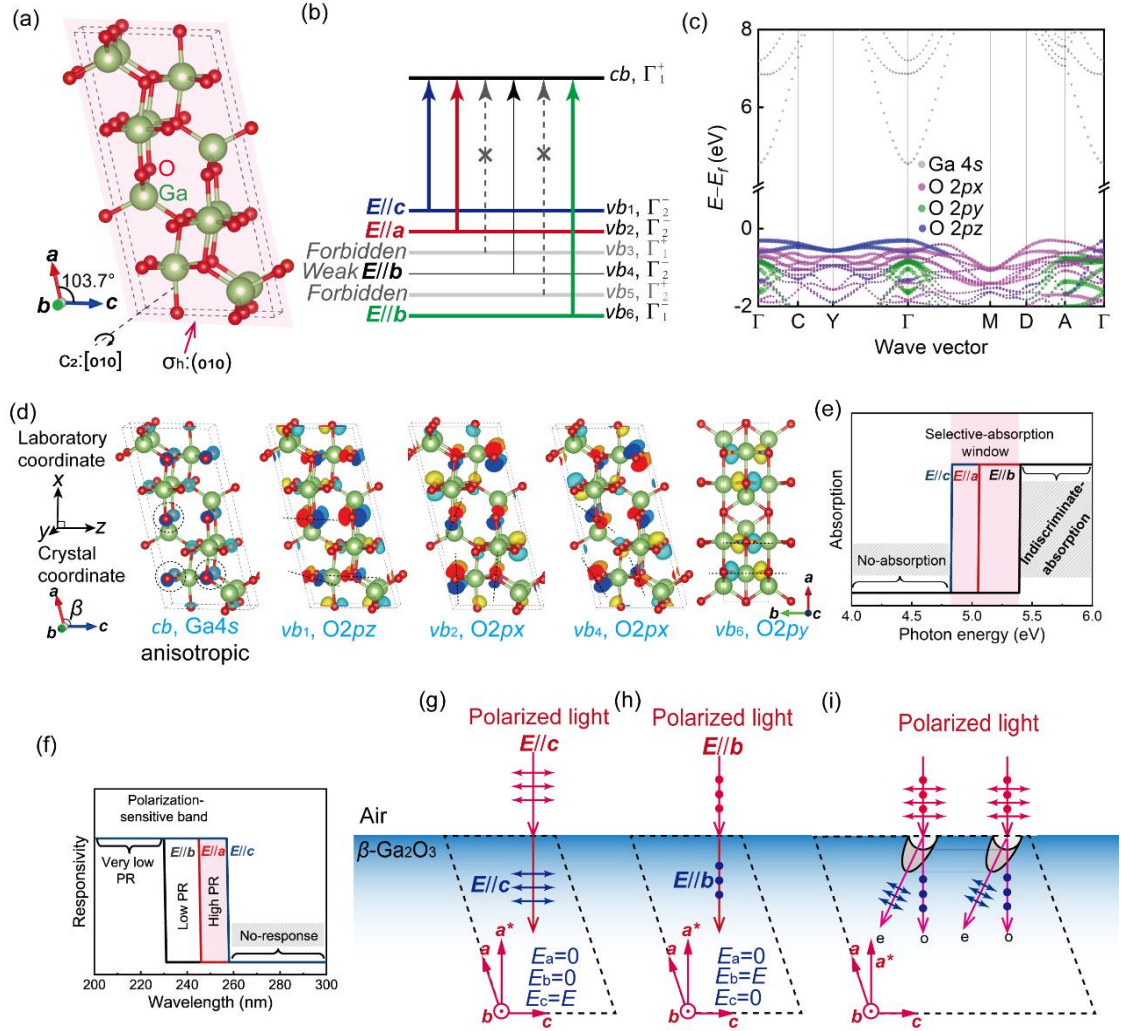

**Figure 2. Underlying mechanism and design principles of *β*-Ga₂O₃ polarization-sensitive photodetector.** (a) Crystal structure of *β*-Ga₂O₃, showing the ***b***-axis two-fold rotation axis and the (010) mirror plane. (b) Band structure at the Γ point and selective optical transitions from the six split valence sub-bands. (c) DFT-calculated orbital-projected band structure of *β*-Ga₂O₃. (d) Electron wave function distributions corresponding to the projected bands $cb$, $vb_1$, $vb_2$, $vb_4$, and $vb_6$. (e) Schematic of polarization-dependent absorption for ***E***//***a***, ***b*** and ***c***. (f) Corresponding photoresponse spectra, highlighting the peak ***E***//***c*** response between 245-258 nm. (g, h) Intracrystal electric field for (g) ***E***//***c*** and (h) ***E***//***b*** polarization under normal incidence on the (100) plane. (i) Electric field decomposition in *β*-Ga₂O₃ when considering birefringence.

## 3. Device performance of *β*-Ga₂O₃ polarization-sensitive photodetector

The *β*-Ga₂O₃ polarization-sensitive photodetector was fabricated by transferring a mechanically exfoliated single-crystal *β*-Ga₂O₃ microribbon onto a quartz substrate, forming a metal–semiconductor–metal (MSM) structure with Au contacts [**Figure 3**(a)]. An optical micrograph of the device [Figure 3(b)] defines a photosensitive area of 3 × 8.9 μm². The thickness of the *β*-Ga₂O₃ layer, determined from atomic force microscopy measurement [AFM, Figure 3(c)], is 460 nm. The microribbon was cleaved along the (100) plane, with its long and short edges aligned with the *b* and *c* crystallographic axes, respectively, as confirmed in our prior work[21].

## 3.1 Performance of *β*-Ga₂O₃ polarization-sensitive photodetector under natural light

As a preliminary characterization of the optoelectronic performance, the current-voltage (*I-V*) characteristics of the device were tested under natural light of 254 nm wavelength. As shown in Figure 3(d), the *β*-Ga₂O₃ detector exhibits an ultra-low $I_{dark}$ of 0.4 pA at a 10 V bias. Consequently, despite a modest photocurrent, a high PDCR is achieved. This characteristic provides a foundation for achieving a high PR in the following tests. The photoresponse of the device under 254 nm illumination and a 10 V bias, measured using periodic DUV cycles (5 s on/off), demonstrates excellent stability and reproducibility. As shown in Figure 3(e), no obvious degradation is observed between the initial and the 200-second cycle (insets). Both the rise and falling times were measured to be less than 20 ms, which is limited by the sampling rate of our source meter. This fast switching and the high stability, underscores the high quality of the single-crystal *β*-Ga₂O₃ material and the robust performance of the fabricated device. The spectral responsivity measurement in Figure 3(f) reveals multiple distinct peaks in the response profile of the *β*-Ga₂O₃ detector. A prominent peak responsivity of 59 A/W is observed at 230 nm, accompanied by another peak of 41.1 A/W around 258 nm. The inset displays the data on a linear scale, where the overall response is schematically decomposed into three individual components centered at approximately 230, 245, and 258 nm[38]. This clear spectral separation of the three distinct absorption edges in *β*-Ga₂O₃ presents a viable pathway to eliminate polarization crosstalk, thereby achieving high PR.

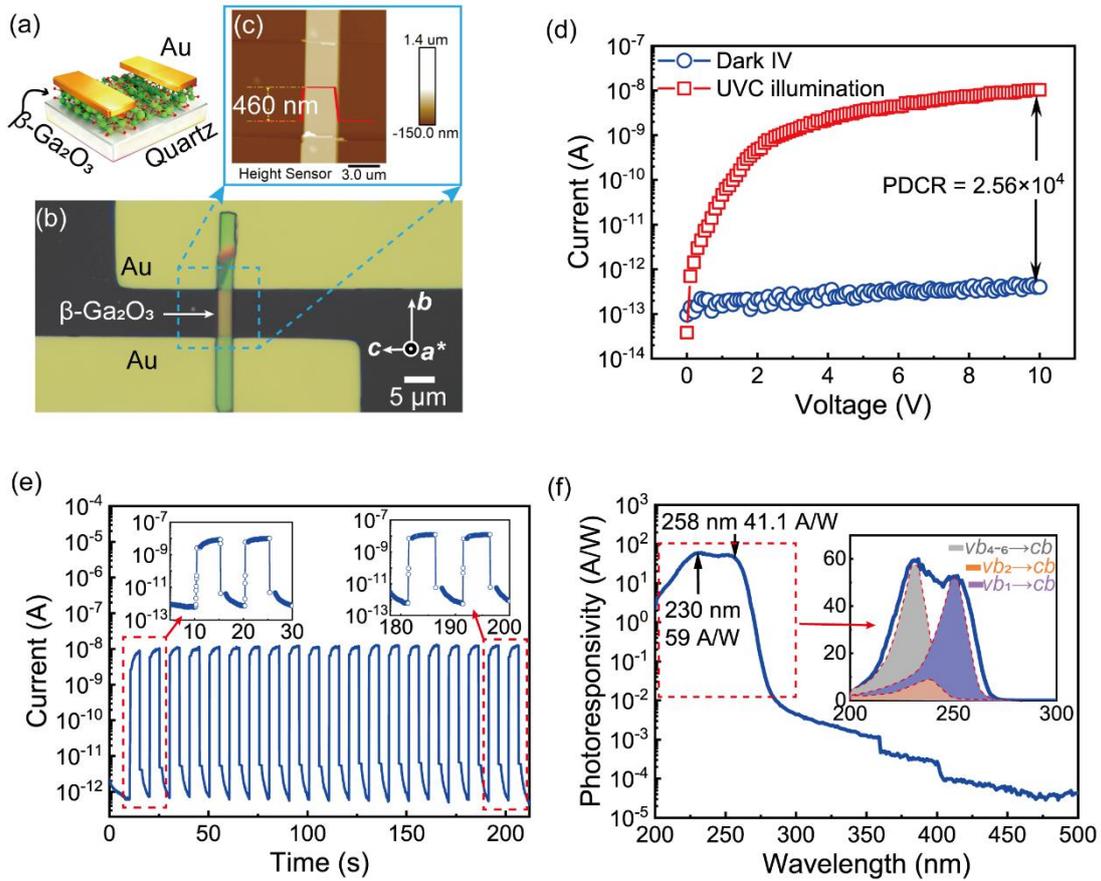

**Figure 3. Device structure and optoelectronic characterizations of the *β*-Ga$_2$O$_3$ photodetector under natural light.** (a) Schematic diagram of the MSM photodetector. (b) Optical microscopy image of the fabricated device. (c) AFM image revealing the thickness of the *β*-Ga$_2$O$_3$ microribbon. (d) *I-V* curves tested under dark and DUV light conditions, showing the ultralow dark current. (e) Time-resolved photoresponse to periodically switched 254 nm illumination at a 10 V bias, demonstrating the device's stability and switching speed. (f) Spectral responsivity plot, showing multiple distinct peaks.

**3.2 Polarization-dependent photoresponse of *β*-Ga$_2$O$_3$ photodetector**

*3.2.1 Photoresponse spectra with varied polarization configurations*

As mentioned in Section 2, the optical properties of the ***b***-axis in *β*-Ga$_2$O$_3$ are decoupled from the ***ac***-plane, with the largest bandgap for ***E*//*b*** and the smallest for ***E*//*c***. Therefore, performing polarization detection in the ***bc***-plane is the optimal choice, as it exhibits the most pronounced difference in optical bandgap between orthogonal polarization directions. As shown in **Figure 4**(a), the optical absorption cutoff edges for ***E*//*b*** (corresponding to a polarization angle $\theta = 90°$) and ***E*//*c*** ($\theta = 0°$) differ by approximately 20 nm. This difference is directly reflected in the photocurrent spectra in Figure 4(b), the photocurrent excited by *E*//*b* -polarized light cuts off at 262 nm (falling below the noise level), while the cutoff edge for *E*//*c* -polarized light red-shifts to 282 nm. Further analysis of Figure 4(b) reveals that as the polarization direction changes, the main peak of the photocurrent spectrum gradually shifts from 233 nm to 258 nm. This indicates that the dominant carrier generation mechanism changes from transition from $vb_1$ band to CBM to transitions from $vb_6$ bands to CBM as $\theta$ increases. Correspondingly, the polarization-dependent responsivity spectra [Figure 4(c)] follow the same trend: the peak responsivity evolves from 233 nm (32 A/W) under *E*//*b* polarization to 258 nm (73 A/W) under *E*//*c* polarization. The inset shows that the PR value reaches a maximum of 547 at 265 nm and drops to a minimum of 0.05 at 230 nm. This reversal of PR is a direct manifestation of the competition between these two transition mechanisms with distinct polarization dependencies. From the polar plots in Figure 4(d), it can be intuitively observed that the periodicity of the photoresponsivity differs markedly at different wavelengths. At 233 nm, where $vb_{4-6}$ to CBM transition dominates, the responsivity follows a $\sin^2\theta$ dependence, being maximum at $\theta = 90°$ and minimum at $\theta = 0°$. At 258 nm, where $vb_1$ band to CBM transition becomes dominant, the responsivity approximately follows a $\cos^2\theta$ dependence, being minimum at $\theta = 90°$ and maximum at $\theta = 0°$. These two transition mechanisms exhibit an orthogonal, complementary relationship. At the intermediate wavelength of 245 nm, their contributions counteract each other, resulting in the poorest polarization selectivity, making it unsuitable for polarization detection. Therefore, in practical applications, wavelengths with well-defined polarization characteristics, such as 258 nm (highest responsivity) or 265 nm (maximum PR)—should be prioritized as operating wavelengths.

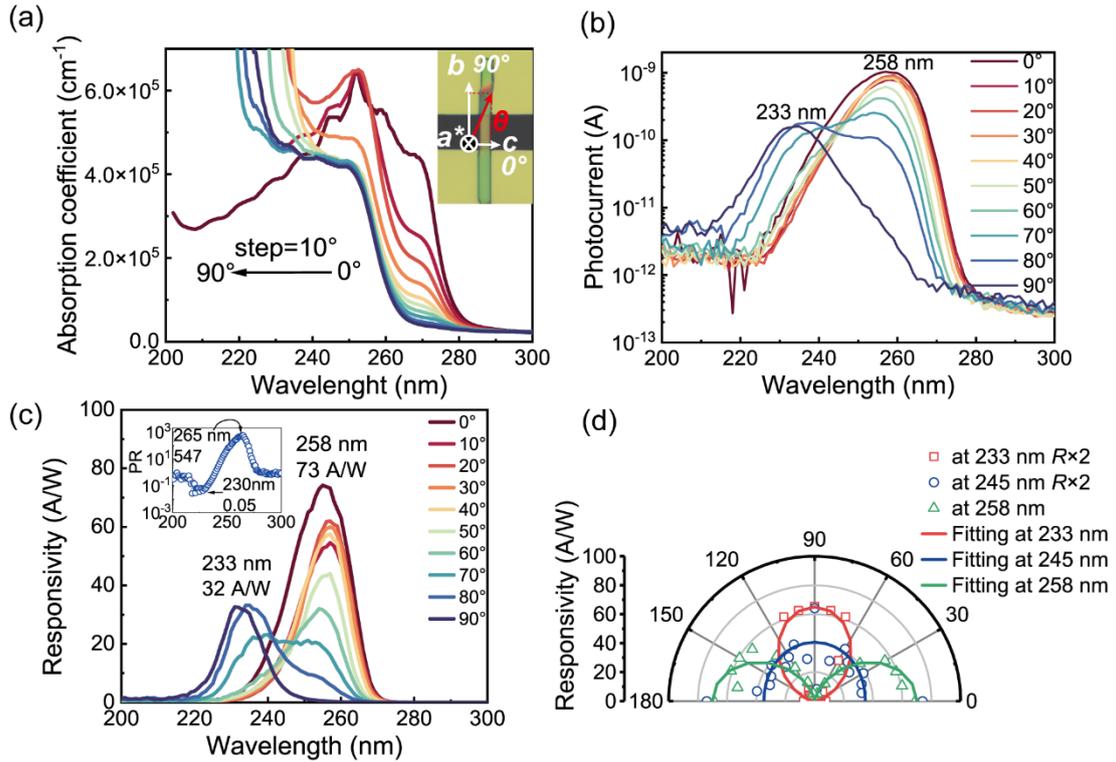

**Figure 4. Polarization-dependent photoresponse of *β*-Ga₂O₃ photodetector.** (a) Absorption spectra of *β*-Ga₂O₃ under linearly polarized light of different angles. The inset defines the polarization angle $\theta$. (b) Photocurrent spectra and (c) responsivity spectra of the *β*-Ga₂O₃ polarization-sensitive photodetector under polarized light from $\theta = 0°$ to $90°$. The inset of (c) shows the wavelength dependence of the PR. (d) Polar plots of the responsivity at 233, 245, and 258 nm, highlighting their distinct polarization anisotropy. The values at 233 and 245 nm are multiplied by 2 for clarity.

*3.2.2 I-V and response speed characterizations with varied polarization angles*

The *I-V* characteristics under polarized illumination at 233 nm and 258 nm are presented in **Figure 5**(a) and (b), respectively. A distinct polarization reversal is observed in the *I-V* characteristics between the two specific wavelengths. As shown in Figure 5(a), illumination at 233 nm generates the strongest photocurrent under ***E***//***b*** polarization ($\theta = 90°$) and the weakest under ***E***//***c*** polarization ($\theta = 0°/180°$). The trend is completely reversed at 258 nm [Figure 5(b)], where the photocurrent is maximized under ***E***//***c*** polarization and minimized under ***E***//***b*** polarization. The time-resolved photoresponse at 258 nm exhibits a clear $\cos^2\theta$ dependence on polarization angle [Figure 5(c)], which originates from the selective excitation of the *vb₁*-to-CBM transition [Figure 5(d)]. For this transition, only the electric field component along the

*c*-axis ($E\cos\theta$) is effective. The orthogonal component ($E\sin\theta$) corresponds to higher-energy transitions (e.g., from $vb_6$ band) that cannot be activated at this photon energy. The higher photon energy at 233 nm enables both transition pathways [Figure 5(f)], leading to a photoresponse described by ($\cos^2\theta+20\sin^2\theta$) [Figure 5(e)]. In this regime, the $E\sin\theta$ component drives transitions from the $vb_6$ band, while the $E\cos\theta$ component concurrently excites the $vb_1$-to-CBM transition in the vicinity of the Γ-point. The observed photocurrent is thus a weighted sum of $\sin^2\theta$ and $\cos^2\theta$ terms, quantified by equation (5).

$$I \propto (R_b \sin^2\theta + R_c \cos^2\theta) \tag{5}$$

where $R_b$ and $R_c$ are photoresponsivity for 90° and 0° polarized light. Over a half polarization cycle (0°–90°), the photoresponse exhibits opposite trends: growing with $\theta$ at 233 nm but diminishing at 258 nm [Figure 5(g) and (h)]. Normalized dynamics reveal fully isotropic rise/fall times at 258 nm (~3.27/0.28 s), whereas at 233 nm, while the rise time is isotropic (~4 s), the fall time is anisotropic, lengthening from ~0.24 s (90°) to ~1.15 s (0°) [Figure 5(i) and (j)]. These contrasting polarization dependencies are summarized in the polar plot of Figure 5(k): a near-circular profile for 258 nm versus a directionally dependent fall time for 233 nm. The observed contrast points to different defect-mediated transition and trapping processes at 258 nm and 233 nm. Elucidating the precise defect identities remains for future work, as the present study centers on reporting the intrinsic band-to-band (VB-to-CBM) transition.

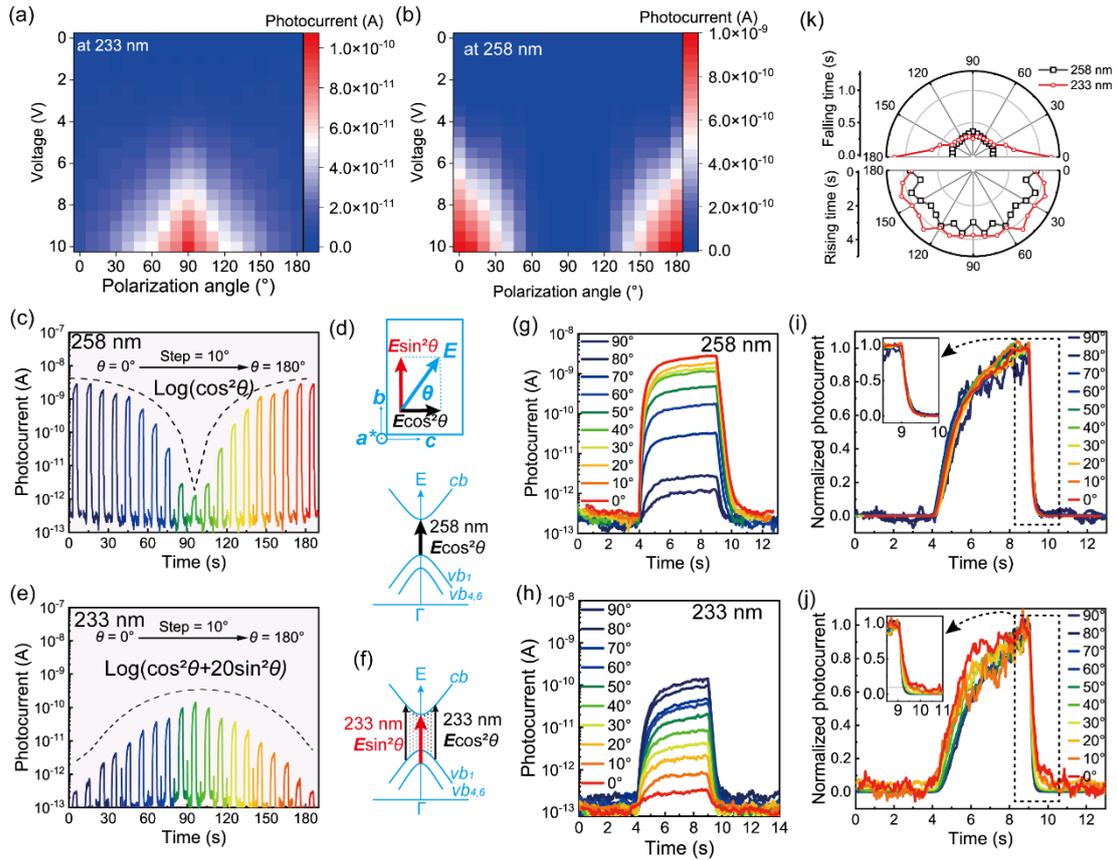

**Figure 5. Polarization-dependent photoresponse tests.** (a, b) *I-V* characteristics under polarized illumination at 233nm and 258 nm, respectively. (c, e) Time-resolved photoresponse at (c) 258 nm and (e) 233 nm. (d, f) Schematic diagrams of transition processes for (d) 258 nm and (f) 233 nm illumination with varied polarization directions. (g, h) A magnified view of a single response cycle at (g) 258 nm and (h) 233 nm for $\theta$ from 0° to 90°. (i, j) Corresponding normalized temporal responses at (i) 258 nm and (j) 233 nm. Insets are the enlarged view of the falling edges. (k) Polar plots summarizing the rise and falling times at both wavelengths across varied polarization angles.

*3.2.3 Applications on PDM communication and Stokes vector retrieval*

To demonstrate its practical utility, the *β*-Ga$_2$O$_3$ polarization-sensitive photodetector was applied to a PDM free-space communication link in DUV band. **Figure 6**(a) illustrates the PDM communication process: the transmitter (Alice) modulates a 233 nm light source using an encoder composed of linear polarizers at 0°, 45°, and 90°, along with a complete shutter, thereby encoding information into distinct polarization states and intensities. The specific encoding scheme is detailed in the inset table. The binary logic for signal decoding is defined as follows: the code "00" is assigned when neither the horizontal nor the vertical analyzer at the receiver (Bob) detects a signal; "10" when only the horizontal analyzer registers a signal; "01" when only the vertical analyzer does; and "11" when both analyzers detect the signal. As a concrete example, Alice transmits a stream of optical signals through the PDM channel (encoded as: dark, 45°, dark, 45°, dark, 90°, 0°, 45°, dark, 45°, dark, 90°, dark, 45°, dark, dark). As shown in Figure 6(b), the horizontal analyzer at Bob's side receives the signal sequence 0 1 0 1 0 0 1 1 0 1 0 0 0 1 0 0, while the vertical analyzer receives 0 1 0 1 0 1 0 1 0 1 0 1 0 1 0 0. Through simple ASCII decoding, Bob can recover the message "SD" and "UT" sent by Alice. The advantage of this PDM scheme is that it doubles the information transmission capacity without requiring an additional physical channel. Furthermore, this mechanism inherently enhances data security, as any eavesdropper (Eve) lacking the specific analyzer basis configuration gains no access to the transmitted information. In this specific demonstration, since the encoding is restricted to a finite set of four bases (dark, 0°, 45°, 90°), the transmitted code can be straightforwardly retrieved by Bob. For the general case of arbitrary polarization states, complete information retrieval requires solving for the Stokes parameters, which the polarization-sensitive photodetector also enables.

Following the established principle of division-of-focal-plane (DoFP) polarization detection, a $\beta$-Ga$_2$O$_3$ polarimetry pixel can be composed of four sub-pixels oriented at 0°, 45°, 90°, and 135°. As illustrated in Figure 6(d), ideally, for incident linearly polarized light with an azimuth angle $\theta$, the photocurrent $I_{0°}$ from the 0°-oriented sub-pixel would ideally be excited solely by the projected component of the electric field $E$ along its axis, i.e., $E\cos\theta$. In practice, however, polarization crosstalk exists, causing $I_{0°}$ to also receive a contribution from the orthogonal field component $E\sin\theta$. Consequently, the total photocurrent is a sum of both contributions. The same principle applies to the sub-pixels at 45°, 90° and 135°, where each total photocurrent comprises a "signal" component from its primary orientation and a "crosstalk" component from the orthogonal direction. Conversion from photocurrent to light intensity is required to calculate the Stokes parameters ($S_0$, $S_1$, $S_2$). For an ideal polarizer with negligible crosstalk, this is simply achieved by dividing each sub-pixel's photocurrent by its responsivity along the primary detection axis. In devices with non-negligible polarization crosstalk, however, this direct conversion introduces systematic error. This is the fundamental error source in polarimetry caused by crosstalk. A key advancement in this work is the significant suppression of polarization crosstalk through optimized detection wavelength, crystal plane and angle of incidence. Leveraging this optimization, the response of each sub-pixel can be justifiably approximated as dominated by its primary orientation. This allows us to accurately convert photocurrents to intensities using a unified responsivity $R_c$, and subsequently retrieve the Stokes parameters with high fidelity.

As a demonstration case, the $\beta$-Ga$_2$O$_3$ polarimeter device was sequentially placed at four distinct orientations to emulate a DoFP pixel illuminated by linearly polarized light at 75°. The corresponding photocurrents from the four sub-pixels oriented at 0°, 45°, 90°, and 135° (denoted as $I_{0°}$, $I_{45°}$, $I_{90°}$ and $I_{135°}$) are shown in Figure 6(e). Owing to the high polarization ratio of 547 for $\beta$-Ga$_2$O$_3$, the corresponding polarization crosstalk ratio (CR) is as low as 0.18%, a performance level that far exceeding the typical requirement of <1% for coherent PDM communication systems[41,42]. By neglecting the residual crosstalk, these photocurrents were converted into light intensity components along the four directions (0°, 45°, 90°, 135°), as shown in Figure 6(f), and subsequently used to calculate the Stokes parameters $S_0$, $S_1$ and $S_2$ [Figure 6(g)]. $S_0$ representing the total effective optical intensity (i.e., the absorbed light that generates photocurrent), reveals that within the 210-240 nm band, the effective intensity for 0° polarization is substantially higher than that for 90° polarization. In the 245-275 nm band, $S_1$ becomes negative, indicating that the dominant polarization direction for generating a larger photocurrent switched to 90°. This aligns perfectly with the reversal in PR across these two spectral bands. Concurrently, the opposite sign variation trend of $S_2$ compared to $S_1$ indicates that the effective intensity in the 245-275 nm band is primarily contributed by 45° polarization, whereas in the 210-240 nm band, it is dominated by 135° polarization. From these Stokes parameters, the degree of polarization (DoP) and the azimuth angle of polarization (AoP) were calculated as functions of wavelength, with the results plotted in Figure 6(h). It is evident that both DoP and AoP exhibit significant spectral dependence. Notably, within the 265-270 nm band, the DoP approaches 100% while the AoP remains stable at approximately 75° (matching the incident polarization). It demonstrates that polarization detection and Stokes parameter retrieval achieve the highest accuracy within this specific spectral window.

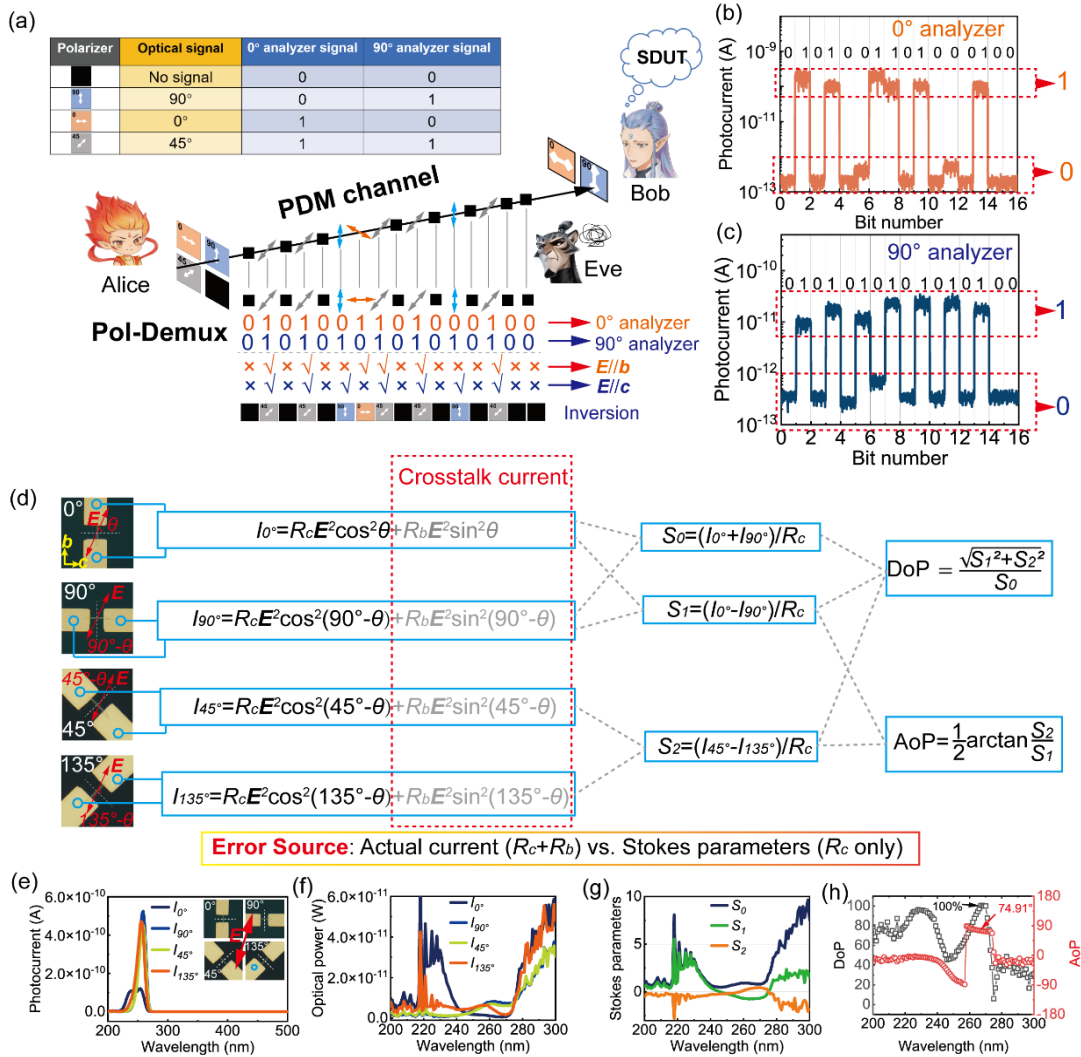

**Figure 6. Schematic of polarization division multiplexing and Stokes vector retrieval using a *β*-Ga₂O₃ polarimeter.** (a) Schematic of the PDM optical communication process. The inset shows the signal encoding truth table. (b, c) Optical response at the 0° (b) and 90° (c) analyzers. (d) The theoretical expressions for the photocurrents from the four sub-pixels in a DoFP pixel, including both the primary signal and polarization crosstalk terms. The method for calculating the Stokes parameters and DoP, AoP derived from them. (e) Photocurrent spectra from the four sub-pixels (0°, 45°, 90°, 135°). (f) Effective optical intensity spectra converted from the photocurrents. (g) Calculated Stokes parameters ($S_0$, $S_1$, $S_2$). (h) Wavelength-dependent

curves of the DoP and the AoP.

## 4. Conclusion

In this work, we systematically demonstrate the design, implementation, and application of $\beta$-Ga$_2$O$_3$-based DUV polarization-sensitive photodetector with inherently suppressed polarization crosstalk. Through a synergistic approach combining group-theoretical analysis, first-principles calculations and experimental validation, we establish a comprehensive framework for achieving high-performance polarization detection. The core of our strategy lies in leveraging the intrinsic material properties of $\beta$-Ga$_2$O$_3$. We identify the critical role of its low-symmetry monoclinic crystal structure (C2/m space group) in enabling pronounced optical anisotropy and well-defined transition selection rules. Specifically, the large valence band splitting energy ($\Delta E_{vb} >$ 0.5 eV) and the symmetry-allowed, yet geometrically orthogonal, interband transitions from $\Gamma_2^-$ and $\Gamma_1^-$ valence sub-bands to the CBM create a spectrally isolated detection window. By operating within the 245–258 nm band with illumination normal to the (100) crystal plane, we exclusively activate the *vb$_1$* ($\Gamma_2^-$)-to-*CBM* transition, which responds only to ***E***//***c*** polarized light. This configuration effectively decouples the response from orthogonal polarizations (***E***//***b***), eliminating the primary source of polarization crosstalk at the physical level. Our fabricated (100)-oriented $\beta$-Ga$_2$O$_3$ single-crystal detector validates this principle, achieving an ultrahigh PR exceeding 547 at 265 nm and an exceptionally low polarization crosstalk ratio of ~0.18%.

We further characterized the device's polarization-resolved performance, confirming a pronounced wavelength-dependent reversal of the dominant polarization response (from ***E***//***b***-dominated at 233 nm to ***E***//***c***-dominated at 258 nm) and its direct correlation with the underlying transition mechanisms. The detector exhibits fast response (~20 ms), excellent stability, and high responsivity (up to 73 A/W at 258 nm). To demonstrate its practical utility, we successfully implemented the $\beta$-Ga$_2$O$_3$ detector in PDM communication and Stokes parameter retrieval applications. In the PDM case, we established a proof-of-concept DUV free-space communication link, where the minimized crosstalk enabled clear encoding/decoding of polarization states, effectively doubling the data transmission capacity without additional channels and enhancing physical-layer security. Furthermore, by emulating a DoFP pixel, we accurately retrieved the full Stokes vector ($S_0$, $S_1$, $S_2$) and derived the DoP and AoP from a single device. Notably, in the 265–270 nm band, a near-100% DoP and stable AoP were achieved, highlighting the spectral window for highest measurement accuracy.

This work shifts the paradigm for DUV polarimetry from reliance on complex external nanostructured filters to the exploitation of intrinsic material anisotropy. *β-Ga$_2$O$_3$* emerges as a premier material platform, offering a direct, robust, and high-performance solution for polarization-sensitive detection. The strategies elucidated here—optimal spectral window selection, crystal orientation engineering, and operating configuration—provide a general blueprint for developing next-generation, high-fidelity polarization photonic devices for advanced optical communication, imaging, and sensing systems.

**Experimental Section**

*Theoretical calculation*: First-principles calculations based on density functional theory were performed using the Vienna Ab initio Simulation Package (VASP). The electron-ion interactions were modeled with the projector augmented-wave (PAW) method. For structural relaxation and electronic structure analysis, the exchange-correlation functional was treated with the Perdew-Burke-Ernzerhof (PBE) formulation within the generalized gradient approximation (GGA). A more accurate description of the electronic band structure was subsequently obtained using the HSE06 hybrid functional, which included 38.4% Hartree-Fock exchange. The plane-wave basis set was expanded with a kinetic energy cutoff of 520 eV, and a Monkhorst-Pack k-point grid of 2 × 8 × 4 was used for sampling the Brillouin zone of the conventional cell. The crystal structure was fully optimized at 0 K, with ionic relaxations proceeding until the forces on all atoms were converged below 0.01 eV/Å. The relaxed bulk *β*-Ga$_2$O$_3$ structure yielded lattice constants of a = 12.40 Å, b = 3.08 Å, and c = 5.70 Å. For the purpose of visualization and subsequent analysis, the crystallographic axes were aligned with the Cartesian laboratory frame (***a**→**x'**, **b**→**y'**, **c\***→**z'***), where ***c\**** represents the direction normal to the ***ab***-plane. To match the experimental coordinate system used in optoelectrical measurements, the computed data were transformed via a rotation matrix $R_y$.

$$R_y = \begin{pmatrix} \cos 13.7° & 0 & \sin 13.7° \\ 0 & 1 & 0 \\ -\sin 13.7° & 0 & \cos 13.7° \end{pmatrix} \quad (6)$$

After this transformation, the final correspondence between the crystallographic and laboratory coordinate systems is ***a*****→*x*, *b*→*y*, *c*→*z***, where ***a*** denotes the normal direction to the ***bc***-plane.

*Material preparation and characterization*: $β$-Ga$_2$O$_3$ microribbons were prepared via mechanical exfoliation from a commercially acquired (-201) unintentionally doped single crystal (purchased from NCT, Japan) using scotch tape. The exfoliated flakes were then transferred onto a quartz substrate via a dry transfer technique. The surface morphology and thickness of the obtained microbelts were characterized by AFM (Bruker Dimension Icon). Polarization-dependent absorption spectra were measured at room temperature using a Hitachi UH4150 UV-Vis-NIR spectrophotometer equipped with a linear polarizer.

*Device fabrication and characterization*: The fabricated devices were constructed directly on the exfoliated $β$-Ga$_2$O$_3$ microribbons transferred onto the quartz substrate. The fabrication process involved standard photolithography patterning, followed by the deposition of gold electrodes via magnetron sputtering. A final lift-off process in acetone defined the final metal contact geometry. The photoelectrical performance of the devices was characterized using a specialized photoresponse measurement system (Zolix Instruments). The optical excitation was provided by a UV-enhanced xenon lamp, with the output wavelength selected and monochromated by a grating spectrometer. Linear polarization of the incident light was achieved using a Glan-Taylor prism (α-BBO crystal, Fujian Crystock, Inc.). The resulting photocurrent was recorded under a constant bias voltage, with electrical signals collected and analyzed by a Keithley 2636B sourcemeter.

**Supporting Information**

The supporting information is available online or from the authors.

**Acknowledgements**

This work was supported by the National Natural Science Foundation of China (Grant No. 12504060, 12574218, 52572167) and Shandong Provincial Natural Science Foundation (Grant No. ZR2025MS1030).

**Conflict of Interest**

The authors declare that they have no conflict of interest.

**Data Availability Statement**

The data that support the findings of this study are available from the corresponding author upon reasonable request.

Received: ((will be filled in by the editorial staff))

Revised: ((will be filled in by the editorial staff))

Published online: ((will be filled in by the editorial staff))

**Table of Contents**

**Giant intrinsic dichroism in *β*-Ga₂O₃ enables filter-free, high-fidelity polarization division multiplexing**

Traditional polarimetry relies on lossy external filters. This work demonstrates that the intrinsic giant dichroism of *β*-Ga₂O₃, arising from its crystal symmetry and large valence band splitting, enables filter-free and high-fidelity polarization detection. By exploiting optical transition selection rules, a DUV photodetector is realized with an ultrahigh polarization ratio and minimal crosstalk, establishing a materials-centric design principle for next-generation polarimetry.

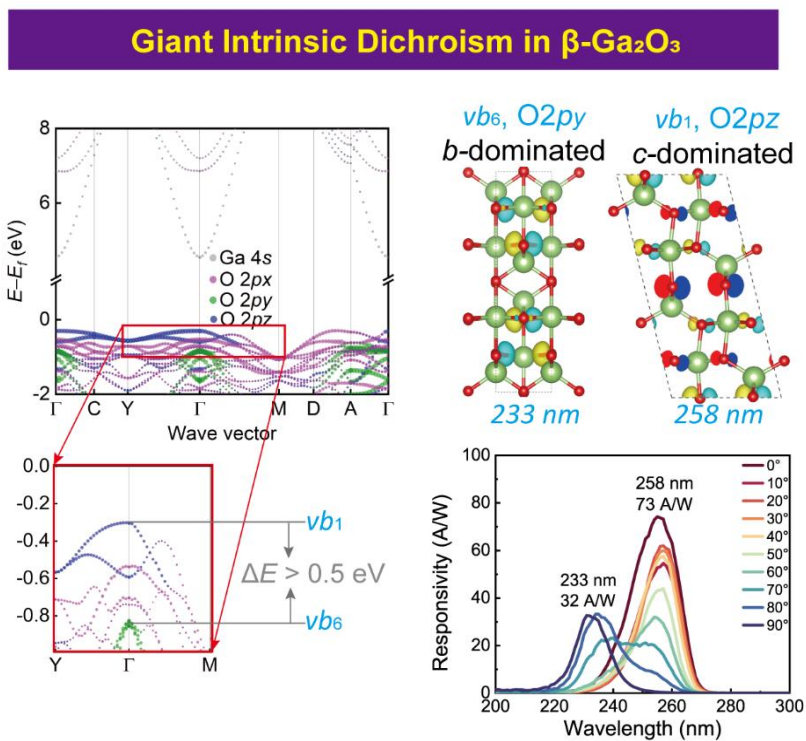